\documentclass[aps,pra,twocolumn,showpacs,superscriptaddress]{revtex4-1}

\usepackage{amsfonts,mathrsfs,amsmath,amsthm,amssymb}
\usepackage{bm,times}
\usepackage{graphicx,epsfig}
\usepackage{ytableau}
\usepackage[colorlinks=true,breaklinks=true,linkcolor=blue,citecolor=blue]{hyperref}

\def \ket {\rangle}
\def \bra {\langle}
\def \tr{ {\rm{Tr}}}
\def \non {\nonumber}

\newcommand{\be}{\begin{eqnarray}}
\newcommand{\ee}{\end{eqnarray}}

\newtheorem{lemma}{Lemma}

\makeatletter
    
    \newcommand{\Rmnum}[1]{\expandafter\@slowromancap\romannumeral #1@}
\makeatother

\begin{document}

\title{Group-theoretic approach for multi-copy programmable discriminators between two unknown qudit states}

\author{Tao Zhou}
\email{zhoutao08@mails.tsinghua.edu.cn}
\affiliation{State Key Laboratory of Low-dimensional Quantum Physics and Department of Physics, Tsinghua University, 100084 Beijing, China}
\affiliation{Department of Physics, Sichuan University, Chengdu, 610064, China}
\author{Xiaohua Wu}
\email{wxhscu@scu.edu.cn}
\affiliation{Department of Physics, Sichuan University, Chengdu, 610064, China}
\author{Gui Lu Long}
\email[Corresponding author: ]{gllong@tsinghua.edu.cn}
\affiliation{State Key Laboratory of Low-dimensional Quantum Physics and Department of Physics, Tsinghua University, 100084 Beijing, China} 
\affiliation{Tsinghua National Laboratory for Information Science and Technology, Tsinghua University, 100084 Beijing, China}

\date{\today}

\begin{abstract}
The discrimination between two unknown states can be performed by a universal programmable discriminator, where the copies of the two possible states are stored in two program systems respectively and the copies of data, which we want to confirm, are provided in the data system. In the present paper, we propose a group-theretic approach to the multi-copy programmable state discrimination problem.  By equivalence of unknown pure states to known mixed states and with the representation theory of $U(n)$ group, we construct the Jordan basis to derive the analytical results for both the optimal unambiguous discrimination and minimum-error discrimination. The POVM operators for unambiguous discrimination and orthogonal measurement operators for minimum-error discrimination are  obtained. We find that the optimal failure probability and minimum-error probability for the discrimination between the mean input mixd states are dependent on the dimension of the unknown qudit states.  We applied the approach to generalize the results of He and Bergou (Phys. Rev. A {\bf 75}, 032316 (2007)) from qubit to qudit case, and we further solve the problem of programmable dicriminators with arbitrary copies of unknown states in both program and data systems.
\end{abstract}

\maketitle

\section{Introduction}
As a recent development, the possibility of discrimination between quantum states can be potentially useful for many applications in quantum communication and quantum computation. In this problem, a quantum state is chosen from a set of known states but we do not know which and want to determine the actual states. This is a nontrivial problem since the states cannot be successfully identified with unit probability because of the non-cloning theorem~\cite{non-cloning}. Two basic strategies have been introduced to achieve the state discrimination, one of which is the minimum-error discrimination~\cite{Helstrom,Holevo,Barnett,Andersson,Chou,Herzog} and the other is the unambiguous discrimination for linearly independent states~\cite{Ivanovic,Dieks,PeresPLA,Jaeger,Chefles,Wu}. In the minimum-error discrimination, errors are permitted and the optimal measurement is required such that the probability of error is minimum, while in the unambiguous discrimination not errors but inconclusive results are permitted, and in the optimal strategy the probability of failure is a minimum. Recently, another approach for the linearly dependent states was proposed with the maximum confidence measurements~\cite{Croke}.

A universal device that can unambiguously discriminate between two unknown qubit states has also been constructed by Bergou and Hillery~\cite{PRL94.160501}. In their work, the system consists of two program qubits $A$ and $C$, and one data qubit $B$. It is assumed that the qubit $A$ and $B$ are prepared in the states $|\psi_1\ket$ and $|\psi_2\ket$ respectively, and qubit $A$ is prepared in either $|\psi_1\ket$ or $|\psi_2\ket$ with probabilities $\eta_1$ and $\eta_2$, where $\eta_1+\eta_2=1$, guaranteeing that the state in system B is always one of the two states. Such a device can measure the total input states
\be
|\Psi_1\ket&=&|\psi_1\ket_A|\psi_1\ket_B|\psi_2\ket_C,\non\\
|\Psi_2\ket&=&|\psi_1\ket_A|\psi_2\ket_B|\psi_2\ket_C,
\ee
where the states $|\psi_1\ket$ and $|\psi_2\ket$ are both unknown,
\be
|\psi_1\ket=a|0\ket+b|1\ket, \ \ \ |\psi_2\ket=c|0\ket+d|1\ket,
\ee
and the parameters $a$, $b$, $c$ and $d$ are all arbitrary unknown complex variables satisfying the normalization conditions $|a|^2+|b|^2=1$ and $|c|^2+|d|^2=1$. This universal discriminator is known as a sort of programmable quantum device, which has been studied in both theory and experiment recently~\cite{PRL79.321,quant0012067,PRA65.022301,PRA66.022112,PRL89.190401,PRA69.032302,Soubusta,D'Ariano}. 

The generalization and the experimental realization aspects of the discriminator above have also been introduced and widely discussed~\cite{PRA72.052306,PRA72.032325,PRA73.012328,PRA73.062334,PLA359.103,PRA72.032310,Zhang,PRA75.032316,Stefan,PRA76.032301,Lucie,PRA78.032320,PRA78.042315,Lin,Sentis,Zhou}. The optimal schemes, where the multiple copies of program and data are used in the input states, have been obtained for $n_A=n_C=n$, $n_B=1$~\cite{PRA72.032325,PRA73.012328,PLA359.103}, for $n_A=n_C=1$, $n_B=n$~\cite{PRA73.062334}, for $n_A=n_C=n,n_B=m$~\cite{PRA75.032316} and for arbitrary copies in both data and program systems~\cite{Sentis}.  The unambiguous discrimination for qudit case has also been considered with single program and data copies ($n_A=n_B=n_C=1$)~\cite{Zhou}.

The most general problem is that there are $n_A$ and $n_C$ copies of states in the program system $A$ and $C$ respectively, and $n_B$ copies of states in the data system $B$, and furthermore, the states are $n$-dimensional ($n\geqslant2$) qudit states rather than qubit states only. Then, the task is to discriminate between two input states,
\be
\label{states}
|\Phi_1\ket&=&|\phi_1\ket_A^{\otimes n_A}|\phi_1\ket_B^{\otimes n_B}|\phi_2\ket_C^{\otimes n_C},\non\\
|\Phi_2\ket&=&|\phi_1\ket_A^{\otimes n_A}|\phi_2\ket_B^{\otimes n_B}|\phi_2\ket_C^{\otimes n_C},
\ee
where $|\phi_1\ket$ and $|\phi_2\ket$ are two unknown states in $n$-dimensional Hilbert space. 

In this paper, we study both the unambiguous discrimination and minimum-error discrimination between two unknown qudit states with the inputs prepared with arbitrary copies in program and data systems as in Eq.~(\ref{states}). Unlike the discrimination between two known states, we cannot only consider this problem in the subspace spanned by the two states $|\phi_1\ket$ and $|\phi_2\ket$, and we should consider it in the full $n$-dimensional space, as the two states are completely unknown to us. By the the equivalence of unknown pure states to known average mixed states as in Refs.~\cite{PRA72.032325,PRA73.012328,PRA73.062334,PRA75.032316,Sentis,Zhou} and with the Jordan-basis method~\cite{PRA73.032107}, we obtain the optimal detection operators and the results for the universal discrimination between the mean states.

The rest of the present paper is organized as follows. Sec.~\ref{sec2} is a preliminary section where we introduce some notations and discuss the average mixed states for the inputs. In Sec.~\ref{sec3}, we will derive the Jordan-basis for the average input states by the reducibility theory of $U(n)$ group. The inner products and their multiplicities are given in Sec.~\ref{sec4} with the coupling theory of angular momenta. The main results of this paper are shown in Sec.~\ref{sec5} and Sec.~\ref{sec6} for optimal unambiguous discrimination and minimum-error discrimination, respectively, and some special examples are discussed in Sec.~\ref{sec7}. Finally, we end this paper with a short summary in Sec.~\ref{sec8}.  Some basic concepts and methods about the group representation theory that are used in this paper are given in the appendix part.

\section{preliminary}\label{sec2} 
In this section, we will discuss the equivalence of unknown pure states to known mixed states. Since the two states $|\phi_1\ket$ and $|\phi_2\ket$ are two unknown states in a $n$-dimensional Hilbert space $\mathcal{H}$, they can change from preparation to preparation. It is only the permutation symmetry properties of $|\Phi_1\ket$ and $|\Phi_2\ket$ that is preserved and can be regarded as available information to distinguish $|\phi_1\ket$ and $|\phi_2\ket$. Therefore, we introduce two density operator
\be
\rho_1&=&\int d\mu(\phi_1)d\mu(\phi_2)[\phi_1^{\otimes n_A}]_A[\phi_1^{\otimes n_B}]_B[\phi_2^{\otimes n_C}]_C,\non\\
\rho_2&=&\int d\mu(\phi_1)d\mu(\phi_2)[\phi_1^{\otimes n_A}]_A[\phi_2^{\otimes n_B}]_B[\phi_2^{\otimes n_C}]_C,
\ee
where $d\mu(\phi)$ is the `natural' measure for the pure state induced by the Haar measure on the unitary group $U(n)$~\cite{measure} with normalization condition $\int d\mu(\phi)=1$. We use $[\phi]$ to denote $|\phi\ket\bra\phi|$ as in the Refs.~\cite{Long,Sentis} and similarly $[\phi\psi\cdots]=[\phi]\otimes[\psi]\otimes\cdots=|\phi\ket\bra\phi|\otimes|\psi\ket\bra\psi|\otimes\cdots$. Without loss of generality, we assume that $n_A\geqslant n_C$, and $n_1=n_A+n_B,n_2=n_B+n_C,N=n_A+n_B+n_C$.
\begin{lemma}
For a pure state $|\psi\ket$ in $n$-dimensional Hilbert space $\mathcal{H}$, 
\be
\int d\mu(\psi)[\psi^{\otimes m}]=\frac{1}{d^{[m]}}\openone^{[m]},
\ee
where $d^{[m]}=\left(\begin{array}{c} n+m-1\\
n-1 \end{array}\right)$ is the dimension of the fully symmetric space $\mathcal{H}^{[m]}$ and $\openone^{[m]}$ is the projector onto this space.
\end{lemma}
{\it Proof: } Since for any vector $|\Psi^{[m]}\ket\in\mathcal{H}^{[m]}$, we have $\int d\mu(\psi)[\psi^{\otimes m}]|\Psi^{[m]}\ket\in\mathcal{H}^{[m]}$, and it is easy to see that $\int d\mu(\psi)[\psi^{\otimes m}]$ satisfies the additivity and scalar multiplication, then it is a linear operator on $\mathcal{H}^{[m]}$. Suppose $U^{[m]}$ is an irreducible representation for $U(n)$ on space $\mathcal{H}^{[m]}$, and therefore
\be
&&U^{[m]}\bigg(\int d\mu(\psi)[\psi^{\otimes m}]\bigg)=\int d\mu(\psi)\big(U|\psi\ket\bra\psi|\big)^{\otimes m}\non\\
&=&\int d\mu(U\psi)[(U|\psi\ket)^{\otimes m}]U^{\otimes m}=\int d\mu(\psi)[\psi^{\otimes m}]U^{\otimes m}\non\\
&=&\bigg(\int d\mu(\psi)[\psi^{\otimes m}]\bigg)U^{[m]},
\ee
where we have used the property $d\mu(U\psi)=d\mu(\psi)$ for Haar measure. According to the Schur's lemma~\cite{Sunbook,Chenbook}, we have $\int d\mu(\psi)[\psi^{\otimes m}]=\lambda\openone^{[m]}$, where $\lambda$ is a constant. Moreover, $\tr\big(\int d\mu(\psi)[\psi^{\otimes m}]\big)=\int d\mu(\psi)\tr([\psi^{\otimes m}])=\int d\mu(\psi)=1$ and $\tr(\openone^{[m]})=d^{[m]}$, so $\lambda=1/d^{[m]}$, which accomplishes the demonstration of the lemma.
\qed

From the lemma above, one can obtain
\be
\label{density}
\rho_1&=&\frac{1}{d_1}\openone^{[n_1]}\otimes\openone^{[n_C]},\non\\
\rho_2&=&\frac{1}{d_2}\openone^{[n_A]}\otimes\openone^{[n_2]},
\ee
where $d_1=d^{[n_1]}d^{[n_C]}$ and $d_2=d^{[n_A]}d^{[n_2]}$ are the ranks for $\rho_1$ and $\rho_2$, respectively. These two mixed states are the ensembles of the two unknown pure states, and the problem becomes the discrimination between $\rho_1$ and $\rho_2$. In the following sections, we will show how to apply the Jordan-basis method to solve this problem.

\section{Jordan basis for the average input states}\label{sec3}
Let us further discuss the structures of mixed states $\rho_1$ and $\rho_2$ in Eq.~(\ref{density}). Consider $\rho_1$ first, and it is obvious that $\openone^{[n_1]}\otimes\openone^{[n_C]}$ is the identity operator on the tensor space $\mathcal{H}^{[n_1]}\otimes\mathcal{H}^{[n_c]}$, whose bases are $\bigg|\begin{array}{c} [n_1] \\ \omega_1\end{array}\bigg\ket\otimes\bigg|\begin{array}{c} [n_C] \\ \omega_C\end{array}\bigg\ket$. Here, the $\omega_1$ and $\omega_C$ correspond to the Weyl tableaux of $[n_1]$ and $[n_C]$, respectively, and we have omitted the label $m_1$ and $m_C$ because they can take only one value. The space $\mathcal{H}^{[n_1]}\otimes\mathcal{H}^{[n_c]}$ is usually reducible under $[U(n)]^{\otimes N}$~\cite{Sunbook,Chenbook}, and the two irreducible bases $\bigg|\begin{array}{c} [n_1] \\ \omega_1\end{array}\bigg\ket$ and $\bigg|\begin{array}{c} [n_C] \\ \omega_C\end{array}\bigg\ket$ can be coupled together to give the irreducible basis~\cite{Chenbook}
\be
\label{irrbasis}
\bigg|\begin{array}{c} [\nu]\\ 
\tau[n_1]m_1[n_C]m_2,\omega\end{array}\bigg\ket=\sum_{\omega_1\omega_C}C^{[\nu]\tau,\omega}_{[n_1]\omega_1,[n_C]\omega_C}\bigg|\begin{array}{c} [n_1] \\ \omega_1\end{array}\bigg\ket\bigg|\begin{array}{c} [n_C] \\ \omega_C\end{array}\bigg\ket,\non\\
\ee
where $C^{[\nu]\tau,\omega}_{[n_1]\omega_1,[n_C]\omega_C}$ are the CG coefficients of the $U(n)$ group, $\omega=1,2,\cdots,d^{[\nu]}$, and $\tau=1,2,\cdots,\{[n_1][n_C][\nu]\}$ is the multiplicity label. According to the Littlewood rule, one has
\be
\label{couple}
[n_1]\otimes[n_C]=\bigoplus_{[\nu]}\{[n_1][n_C][\nu]\}[\nu],
\ee
and it is easy to see $\{[n_1][n_C][\nu]\}=1$ for all possible $[\nu]$ since $[n_1]$ and $[n_C]$ are both totally symmetric. With Young diagrams, Eq.~(\ref{couple}) can be graphically expressed as
\ytableausetup{mathmode, boxsize=1.2em}
\be
\label{coupleY}
&&\overbrace{\begin{ytableau}
{}&{}&\none[\cdots]&{}
\end{ytableau}}^{n_1}\ \otimes\ \overbrace{\ytableausetup{mathmode, boxsize=1.2em}
\begin{ytableau}
{}&{}&\none[\cdots]&{}
\end{ytableau}}^{n_C}=\overbrace{\begin{ytableau}
{}&{}&\none[\cdots]&{}
\end{ytableau}}^N\non\\
&&\oplus\overbrace{\begin{ytableau}
{}&{}&\none[\cdots]&{}\\
{}\end{ytableau}}^{N-1}\oplus\overbrace{\begin{ytableau}
{}&{}&\none[\cdots]&{}\\
{}&{}\end{ytableau}}^{N-2}\oplus\cdots\oplus\overbrace{\begin{ytableau}
{}&\none[\cdots]&{}&{}&\none[\cdots]&{}\\
{}&\none[\cdots]&{}\end{ytableau}}^{n_1}.\non\\
\ee

The Eq.~(\ref{coupleY}) above shows that besides the fully symmetric case, the Young diagram $[\nu]$ can take some special cases with only two rows, such as $[N-1,1],[N-2,2],\cdots,[n_1,n_C]$. The new irreducible basis in Eq.~(\ref{irrbasis}) have two labels $[\nu]$ and $\omega$, and hence can be denoted as $\bigg|\begin{array}{c} [\nu]\\  \omega\end{array}\bigg\ket$ for simplify.  Now, we can see that the basis $\bigg|\begin{array}{c} [\nu]\\  \omega\end{array}\bigg\ket$ also form the complete orthogonal basis for the space $\mathcal{H}^{[n_1]}\otimes\mathcal{H}^{[n_C]}$. Therefore, $\rho_1$ can also be expressed as
\be
\label{Jordan1}
\rho_1=\frac{1}{d_1}\sum_{[\nu],\omega}\bigg|\begin{array}{c} [\nu]\\  \omega\end{array}\bigg\ket\bigg\bra\begin{array}{c} [\nu]\\  \omega\end{array}\bigg|,
\ee
and if $H^{[\nu]}$ is defined as the space spanned by the basis vectors $\bigg|\begin{array}{c} [\nu]\\  \omega\end{array}\bigg\ket$ ($\omega=1,2,\cdots,d^{[\nu]}$), one has
\be
\mathcal{H}^{[n_1]}\otimes\mathcal{H}^{[n_C]}=\bigoplus_{[\nu]}H^{[\nu]},
\ee
where we use $H$ instead of $\mathcal{H}$ because the basis $\bigg|\begin{array}{c} [\nu]\\  \omega\end{array}\bigg\ket$ are not the standard basis of $S_N$. However, $\bigg|\begin{array}{c} [\nu]\\  \omega\end{array}\bigg\ket$ can be transformed to standard basis $\bigg|\begin{array}{c} [\nu]\\ m,\ \omega\end{array}\bigg\ket$ by
\be
\label{nonstan1}
\bigg|\begin{array}{c} [\nu]\\  \omega\end{array}\bigg\ket=\sum_m\bigg\bra \begin{array}{c} [\nu]\\ m\end{array}\bigg|[\nu],\begin{array}{c} [n_1][n_C]\\ m_1m_C\end{array}\bigg\ket\bigg|\begin{array}{c} [\nu]\\ m,\ \omega\end{array}\bigg\ket,
\ee
where $\bigg\bra \begin{array}{c} [\nu]\\ m\end{array}\bigg|[\nu],\begin{array}{c} [n_1][n_C]\\ m_1m_C\end{array}\bigg\ket$ are the $[\nu]\downarrow [n_1]\otimes[n_C]$ subduction coefficients (SDCs) of $S_N$.

Similar discussions can be carried on for $\rho_2$, and
\be
\label{Jordan2}
\rho_2=\frac{1}{d_2}\sum_{[\nu'],\omega'}\bigg|\begin{array}{c} [\nu']\\  \omega'\end{array}\bigg\ket\bigg\bra\begin{array}{c} [\nu']\\  \omega'\end{array}\bigg|,
\ee
where
\be
\label{nonstan2}
\bigg|\begin{array}{c} [\nu']\\  \omega'\end{array}\bigg\ket=\sum_{m'}\bigg\bra \begin{array}{c} [\nu']\\ m'\end{array}\bigg|[\nu'],\begin{array}{c} [n_A][n_2]\\ m_Am_2\end{array}\bigg\ket\bigg|\begin{array}{c} [\nu']\\ m',\ \omega'\end{array}\bigg\ket,
\ee
and
\be
\mathcal{H}^{[n_A]}\otimes\mathcal{H}^{[n_2]}=\bigoplus_{[\nu']}H'^{[\nu']}.
\ee

It should be noticed here that the possible cases for the Young diagram $[\nu']$ are not all the same as those for $[\nu]$ if $n_A\neq n_C$. Since $n_A\geqslant n_C$, $[\nu']$ can take no fewer cases than $[\nu]$, which can be displayed as follows
\ytableausetup{mathmode, boxsize=1.2em}
\be
\label{coupleY2}
&&\overbrace{\begin{ytableau}
{}&{}&\none[\cdots]&{}
\end{ytableau}}^{n_A}\ \otimes\ \overbrace{\ytableausetup{mathmode, boxsize=1.2em}
\begin{ytableau}
{}&{}&\none[\cdots]&{}
\end{ytableau}}^{n_2}=\overbrace{\begin{ytableau}
{}&{}&\none[\cdots]&{}
\end{ytableau}}^N\non\\
&&\oplus\overbrace{\begin{ytableau}
{}&{}&\none[\cdots]&{}\\
{}\end{ytableau}}^{N-1}\oplus\overbrace{\begin{ytableau}
{}&{}&\none[\cdots]&{}\\
{}&{}\end{ytableau}}^{N-2}\oplus\cdots\oplus\overbrace{\begin{ytableau}
{}&\none[\cdots]&{}&{}&\none[\cdots]&{}\\
{}&\none[\cdots]&{}\end{ytableau}}^{\max(n_A,n_2)}.\non\\
\ee
For the last Young diagram on the right hand of the Equation above, the number of the cells for the first row is $\max(n_A,n_2)$. It is easy to see that $\max(n_A,n_2)\leqslant n_1$, and therefore, we can conclude that the Young diagrams $[\nu']$ and $[\nu]$ can take the same possible cases for $n_A=n_C$, while for $n_A>n_C$, $[\nu']$ can take more cases besides those of $[\nu]$. Based on this point, we can also know that $d_1\leqslant d_2$. 

Now, we can introduce some notations that will be useful in the following discussions. Let $\mathbb{H}_{\rm T}=\big(\mathcal{H}^{[n_1]}\otimes\mathcal{H}^{[n_C]}\big)\bigcup\big(\mathcal{H}^{[n_A]}\otimes\mathcal{H}^{[n_2]}\big)$, and this is the total space. For a Young diagram $[\lambda]$ that can be taken for both $[\nu]$ and $[\nu']$, we define $\mathbb{H}=\bigoplus_{[\lambda]}H^{[\lambda]}$, $\mathbb{H}'=\bigoplus_{[\lambda]}H'^{[\lambda]}$, $\mathbb{H}_0=\mathbb{H}\bigcup\mathbb{H'}$, and $\mathbb{H}^{[\lambda]}=H^{[\lambda]}\bigcup H'^{[\lambda]}$. Meanwhile, for a Young diagram $[\mu]$ that $[\nu]$ cannot take but $[\nu']$ can, let $\mathbb{H'}^{\bot}=\bigoplus_{[\mu]}H'^{[\mu]}$. It is evident that these spaces have the following relationships
\be
\mathbb{H}=\mathcal{H}^{[n_1]}\otimes\mathcal{H}^{[n_C]},&&\ \ \mathbb{H'}\oplus\mathbb{H'}^{\bot}=\mathcal{H}^{[n_A]}\otimes\mathcal{H}^{[n_2]}\non,\\
\mathbb{H}_0=\bigoplus_{[\lambda]}\mathbb{H}^{[\lambda]},&&\ \ \mathbb{H}_{\rm T}=\mathbb{H}_0\oplus\mathbb{H'}^{\bot}.
\ee

It has been well known that there exist the Jordan basis for two nonorthogonal Hilbert spaces~\cite{Callagher}, and this can be used to the discrimination between two mixed states if we can find their Jordan bases~\cite{PRA73.032107}. The Jordan bases are defined as follows. The sets of orthogonal and normalized basis $\{|f_1\ket,|f_2\ket,\cdots,|f_k\ket\}$ in space $V_1$ and $\{|g_1\ket,|g_2\ket,\cdots,|g_k\ket\}$ in space $V_2$ form the Jordan bases when
\be
\bra f_i|g_j\ket=\delta_{ij}\cos\theta_i,
\ee
where $\theta_i$ are the so-called Jordan angles $(\theta_1\leqslant\theta_2\leqslant\cdots\leqslant\theta_k)$. Since the support of $\rho_1$ has no overlaps with the space $\mathbb{H'}^{\bot}$, there is no doubt that we should consider the Jordan bases of the two nonorthogonal spaces $\mathbb{H}$ and $\mathbb{H'}$ in subspace $\mathbb{H}_0$. With the Eqs.~(\ref{nonstan1}) and~(\ref{nonstan2}), we obtain the overlap
\begin{widetext}
\be
\label{innerproduct}
\bigg\bra\begin{array}{c} [\nu]\\ \omega\end{array}\bigg|\begin{array}{c} [\nu']\\  \omega'\end{array}\bigg\ket&=&\sum_{m,m'}\bigg\bra \begin{array}{c} [\nu]\\ m\end{array}\bigg|[\nu],\begin{array}{c} [n_1][n_C]\\ m_1m_C\end{array}\bigg\ket\bigg\bra \begin{array}{c} [\nu']\\ m'\end{array}\bigg|[\nu'],\begin{array}{c} [n_A][n_2]\\ m_Am_2\end{array}\bigg\ket\bigg\bra\begin{array}{c} [\nu]\\m,\  \omega\end{array}\bigg|\begin{array}{c} [\nu']\\m',\ \omega'\end{array}\bigg\ket\non\\
&=&\sum_{m}\bigg\bra \begin{array}{c} [\nu]\\ m\end{array}\bigg|[\nu],\begin{array}{c} [n_1][n_C]\\ m_1m_C\end{array}\bigg\ket\bigg\bra \begin{array}{c} [\nu']\\ m\end{array}\bigg|[\nu'],\begin{array}{c} [n_A][n_2]\\ m_Am_2\end{array}\bigg\ket\delta_{[\nu][\nu']}\delta{\omega\omega'},
\ee
\end{widetext}
which shows that  $\bigg|\begin{array}{c} [\nu]\\  \omega\end{array}\bigg\ket$ and  $\bigg|\begin{array}{c} [\nu']\\  \omega'\end{array}\bigg\ket$ have already been the Jordan bases of subspaces $\mathbb{H}$ and $\mathbb{H'}$ . Next, we will consider how to calculate the inner products of Jordan bases and give their multiplicities.

\section{Inner products of Joran basis and the multiplicities}\label{sec4}
Eq.~(\ref{innerproduct}) has shown that the inner products $\bigg\bra\begin{array}{c} [\nu]\\ \omega\end{array}\bigg|\begin{array}{c} [\nu']\\  \omega'\end{array}\bigg\ket$ are dependent on the SDCs only. If we can calculate the SDCs, the inner products can easily be obtained. As we know, it is not a simple work to calculate the SDCs in this paper. However, in another way, the SDCs are connected with the permutation group only, and they are independent on the dimension of $\mathcal{H}$. This suggests that the inner products for any $n$ can be solved as soon as one can calculate those for a special $n$. Fortunately, for $n=2$, the qubit cases, the irreducible basis are just the angular momentum basis of the total system cosisting of A, B and C, and therefore, the inner products can be calculated by the coupling theory of angular momentum, without calculating the exact values of SDCs.

For $n=2$, each copy is a qubit state, and then can been viewed as a spin-$1/2$ system with $j=1/2$. The program systems A and C, and data system B can be regarded as angular systems with quantum numbers $j_A=n_A/2$, $j_C=n_C/2$ and $j_B=n_B/2$. The angular momentum basis $|(j_Aj_B)j_{AB},j_C;JM\ket$ and $|j_A,(j_Bj_C)j_{BC};J'M'\ket$ are the irreducible basis for $\rho_1$ and $\rho_2$, respectively. There is a a one-to-one relationship between the quantum numbers $J$ or $J'$ and the possible Young diagrams in the right hand of Eq.~(\ref{coupleY}) or~(\ref{coupleY2}), and then the quantum numbers $J$ for $\rho_1$ can take no more values than $J'$ for $\rho_2$. In $\rho_1$ the first $n_1$ spins are couple in a symmetric way, therefore $j_{AB}=n_1/2$, and similarly $j_{BC}=n_2/2$ for the same reason. So we come to that $J=(n_1-n_C)/2,\cdots,N/2$ and $J'=|n_A-n_2|/2,\cdots,N/2$. Now,
\be
&&\bra(j_Aj_B)j_{AB},j_C;JM|j_A,(j_Bj_C)j_{BC};J'M'\ket\non\\
&=&(-1)^{j_A+j_B+j_C+J}\sqrt{(2j_{AB}+1)(2j_{BC}+1)}\non\\
&&\times\left\{\begin{array}{ccc}j_A & j_B & j_{AB}\\
j_C & J & j_{BC}\end{array}\right\}\delta_{JJ'}\delta_{MM'}\non\\
&=&\delta_{JJ'}\delta_{MM'}\sqrt{\frac{\left(\begin{array}{c} n_1-k\\n_B\end{array}\right)\left(\begin{array}{c} n_2-k\\n_B\end{array}\right)}{\left(\begin{array}{c} n_1\\n_B\end{array}\right)\left(\begin{array}{c} n_2\\n_B\end{array}\right)}}
\ee
where we have set $J=N/2-k$ ($k=0,1,\cdots,n_C$) and $\left\{\begin{array}{ccc}j_A & j_B & j_{AB}\\
j_C & J & j_{BC}\end{array}\right\}$ are the Wigner's $6j$ symbols~\cite{Zengbook,Edmonds}. The overlaps are independent of the quantum number $M$, and therefore, the inner products of Jordan basis $\bigg\bra\begin{array}{c} [\nu]\\ \omega\end{array}\bigg|\begin{array}{c} [\nu']\\  \omega'\end{array}\bigg\ket$ are determined only by the Young diagram, which means that the inner products are the invariants of $U(n)$ group. For an inner product corresponding to a Young diagram $[\lambda]$ (this diagram can be taken by both $[\nu]$ for $\rho_1$ and $[\nu']$ for $\rho_2$), the multiplicity is the dimension $d^{[\lambda]}$, the number of values that $\omega$ can takes. The inner products of Jordan basis and their multiplicities are all listed in the following table:
\begin{widetext}
\begin{center}
\begin{tabular}{c|c|c|c}
$J\ (n=2)$ & Young diagrams $[\lambda]$ & Inner product $O_k=O^{[\lambda]}$ & Multiplicity $d^k=d^{[\lambda]}$\\
\hline $\frac{N}{2}$ & $[N]$ & 1 & $\left(\begin{array}{c}N+n-1\\ n-1\end{array}\right)$\\
$\frac{N}{2}-1$ &$[N-1,1]$ & $\sqrt{\frac{n_An_C}{n_1n_2}}$ & $\frac{(N-1)(n-1)}{N}\left(\begin{array}{c}N+n-2\\ n-1\end{array}\right)$\\
$\vdots$ & $\vdots$ & $\vdots$ & $\vdots$\\
$\frac{N}{2}-k$ &$[N-k,k]$ & $\sqrt{\frac{\left(\begin{array}{c} n_1-k\\n_B\end{array}\right)\left(\begin{array}{c} n_2-k\\n_B\end{array}\right)}{\left(\begin{array}{c} n_1\\n_B\end{array}\right)\left(\begin{array}{c} n_2\\n_B\end{array}\right)}}$ & $\frac{N-2k+1}{N-k+1}\left(\begin{array}{c}N+n-k-1\\ n-1\end{array}\right)\left(\begin{array}{c}n+k-2\\ n-2\end{array}\right)$\\
$\vdots$ &$\vdots$ & $\vdots$ & $\vdots$\\
$\frac{N}{2}-n_C$ &$[n_1,n_C]$ & $\sqrt{\frac{n_A!n_B!n_C!(n_1-n_C)!}{n_1!n_2!(n_A-n_C)!}}$ & $\frac{n_1-n_C+1}{n_1+1}\left(\begin{array}{c}n_1+n-1\\ n-1\end{array}\right)\left(\begin{array}{c}n+n_C-2\\ n-2\end{array}\right)$\\
\end{tabular}
\end{center}
\end{widetext}
For the qubit case ($n=2$), the inner products and multiplicities will reduce to those in Ref.~\cite{PRA75.032316}, if we further assume $n_A=n_C$. In that paper, the authors found an inherent symmetry to study the structures of the mean input states, which works only for $n_A=n_C$.

With the Jordan basis above, the subspace $\mathbb{H}^{[\lambda]}$ can be further decomposed into $\mathbb{H}^{[\lambda]}=\bigoplus_\omega \mathbb{H}^{[\lambda]}_\omega$, where $\mathbb{H}^{[\lambda]}_\omega$ is the subspace spanned by $\bigg|\begin{array}{c} [\lambda]\\  \omega\end{array}\bigg\ket_1$ and  $\bigg|\begin{array}{c} [\lambda]\\  \omega\end{array}\bigg\ket_2$, $\omega=1,2,\cdots,d^{[\lambda]}$, and the indexes ``$1$'' and ``$2$'' are used to label the bases for $\rho_1$ and $\rho_2$, respectively. In the overall ensemble, since $\rho_1$ occurs with probability $\eta_1$ and $\rho_2$ with $\eta_2$, the probability of occurrence for $\bigg|\begin{array}{c} [\lambda]\\  \omega\end{array}\bigg\ket_1$ is $\eta_1/d_1$ and that of $\bigg|\begin{array}{c} [\lambda]\\  \omega\end{array}\bigg\ket_2$ is $\eta_2/d_2$. Therefore, the probability for the occurrence of a vector in $\mathbb{H}^{[\lambda]}_\omega$ is
\be
p^{[\lambda]}_\omega=\frac{\eta_1}{d_1}+\frac{\eta_2}{d_2}.
\ee
The probability that $\bigg|\begin{array}{c} [\lambda]\\  \omega\end{array}\bigg\ket_1$ occurs conditioned on that $\mathbb{H}^{[\lambda]}_\omega$ has occurred is $\eta^{[\lambda]}_{\omega,1}=\eta_1/(d_1p^{[\lambda]}_\omega)$ and similarly $\eta^{[\lambda]}_{\omega,2}=\eta_2/(d_2p^{[\lambda]}_\omega)$. Finally, the problem to discriminate between $\rho_1$ and $\rho_2$ is reduced to deriving the optimal schemes for the unambiguous and the minimum-error discrimination between two pure states occurring with probabilities $\eta^{[\lambda]}_{\omega,1}$ and $\eta^{[\lambda]}_{\omega,2}$ in each subspace $\mathbb{H}^{[\lambda]}_\omega$.

\section{Optimal unambiguous discrimination}\label{sec5}
To discriminate between $\bigg|\begin{array}{c} [\lambda]\\  \omega\end{array}\bigg\ket_1$ and $\bigg|\begin{array}{c} [\lambda]\\  \omega\end{array}\bigg\ket_2$ with the  ${\sl a~ priori}$ probabilities $\eta^{[\lambda]}_{\omega,1}$ and $\eta^{[\lambda]}_{\omega,2}$ in space $\mathbb{H}^{[H]}_\omega$ unambiguously, we can introduce the POVM operators in the following form~\cite{PRA73.032107,PRA75.032316}
\be
\Pi^k_{\omega,1}(q_{k,1},q_{k,2})&=&\frac{1-q_{k,1}}{1-O_k^2}|\psi_{k,\omega}^\bot\ket_2\bra\psi_{k,\omega}^\bot|,\non\\
\Pi^k_{\omega,2}(q_{k,1},q_{k,2})&=&\frac{1-q_{k,2}}{1-O_k^2}|\psi_{k,\omega}^\bot\ket_1\bra\psi_{k,\omega}^\bot|,\non\\
\Pi^k_{\omega,0}(q_{k,1},q_{k,2})&=&\openone^k_\omega-\Pi^k_{\omega,1}-\Pi^k_{\omega,2}
\ee
where we have used $k$ to denote the Young diagrams listed in the table in Sec.~\ref{sec4}. $q_{k,1}$  or $q_{k,2}$ is the failure probability for $\bigg|\begin{array}{c} [\lambda]\\  \omega\end{array}\bigg\ket_1$ or $\bigg|\begin{array}{c} [\lambda]\\  \omega\end{array}\bigg\ket_2$ in the unambiguous discrimination, the normalized vector $|\psi_{k,\omega}^\bot\ket_{1(2)}$ is orthogonal to $\bigg|\begin{array}{c} [\lambda]\\  \omega\end{array}\bigg\ket_{1(2)}$ in the subspace $\mathbb{H}^{[\lambda]}_\omega$, and $\openone^k_\omega$ is the unit operator in $\mathbb{H}^{[\lambda]}_\omega$. The parameters $q_{k,1}$ and $q_{k,2}$ are independent of $\omega$ because the inner products are independent of $\omega$. The total failure probability for the unambiguous discrimination between $\rho_1$ and $\rho_2$ is
\be
Q=\sum_{k=0}^{n_C}d^kQ_k=\sum_{k=0}^{n_C}d^k(\frac{\eta_1q_{k,1}}{d_1}+\frac{\eta_2q_{k,2}}{d_2}),
\ee
with $d^k=d^{[N-k,k]}$ the multiplicities for the inner products of Jordan basis. We can find the the optimal settings
\be
q_{k,1}^{\rm opt}=\left\{\begin{array}{ll} 1 & \textrm{if } \eta_1<c_k\\
\sqrt{\frac{\eta_2d_1}{\eta_1d_2}}O_k & \textrm{if } c_k\leqslant\eta_1\leqslant d_k\\
O_k & \textrm{if } \eta_1\geqslant d_k\end{array}\right.,
\ee
and $q_{k,2}^{\rm opt}=O_k^2/q_{k,1}^{\rm opt}$, where $Q_k$ attains its minimum,
\be
Q_k^{\rm opt}=\left\{\begin{array}{ll} \frac{\eta_1}{d_1}+\frac{\eta_2}{d_2}O_k^2 & \textrm{if } \eta_1<c_k\\
2\sqrt{\frac{\eta_1\eta_2}{d_1d_2}}O_k & \textrm{if } c_k\leqslant\eta_1\leqslant d_k\\
\frac{\eta_1}{d_1}O_k^2+\frac{\eta_2}{d_2} & \textrm{if } \eta_1\geqslant d_k\end{array}\right..
\ee
The boundaries $c_k$ and $d_k$ are as follows
\be
c_k=\frac{d_1O_k^2}{d_2+d_1O_k^2}, \ d_k=\frac{d_1}{d_1+d_2O_k^2}.
\ee

Finally, the optimal failure probability for the unambiguous discrimination between $\rho_1$ and $\rho_2$ is 
\be
\label{UDP}
Q^{\rm opt}=\sum_{k=0}^{n_C}d^kQ_k^{\rm opt}
\ee
and the corresponding optimal POVM are
\be
\label{POVM}
\Pi_1&=&\sum_{k,\omega}\Pi^k_{\omega,1}(q_{k,1}^{\rm opt},q_{k,2}^{\rm opt}),\non\\
\Pi_2&=&\sum_{k,\omega}\Pi^k_{\omega,2}(q_{k,1}^{\rm opt},q_{k,2}^{\rm opt})+\openone^\bot,\non\\
\Pi_0&=&\openone_{\rm T}-\Pi_1-\Pi_2,
\ee
where $\openone_{\rm T}$ is the identity operator on the space $\mathbb{H}_{\rm T}$, and $\openone^\bot$ is the projector onto the subspace $\mathbb{H}^\bot$. The projector $\openone^\bot$ appears in $\Pi_2$ because the occurrence in $\mathbb{H}^\bot$ always means the input state is $\rho_2$ (or $|\Phi_2\ket$). We see from the equations above that both the POVM operators and the optimal failure probability of unambiguous discrimination between $\rho_1$ and $\rho_2$ are dependent on the dimension $n$ and the numbers of copies in data system and program systems, since the parameters such as $O_k$ and $d^k$ are dependent on them.

\section{Minimum-error discrimination}\label{sec6}
For the minimum-error discrimination between the two mixed states $\rho_1$ and $\rho_2$, the inconclusive results do not occur, so $\Pi_0=0$, and we require that the probability of errors in the discrimination procedure is a minimum. The error probability can be expressed as~\cite{Helstrom}
\be
P_E=\eta_1\tr(\rho_1\Pi_2)+\eta_2\tr(\rho_2\Pi_1)=\eta_1+\tr(\Lambda\Pi_1),
\ee
where $\Lambda=\eta_2\rho_2-\eta_1\rho_1=\sum_i\lambda_i|\omega_i\ket\bra\omega_i|$, with $\lambda_i$ the eigenvalue spectrum of the operator $\Lambda$. It is obvious that the minimum of the error probability is obtained when $\Pi_1$ is the projector onto the space spanned by those eigenstates $|\omega_i\ket$ that belong to negative eigenvalues $\lambda_i$. The optimal detection operators therefore read
\be
\Pi_1=\sum_{i<i_0}|\omega_i\ket\bra\omega_i|,\ \ \Pi_2=\sum_{i\geqslant i_0}|\omega_i\ket\bra\omega_i|,
\ee
where $\omega_i<0$ for $1\leqslant i\leqslant i_0$ and $\omega_i\geqslant0$ for $i\geqslant i_0$. Clearly, the minimum-error measurement for discriminating between two quantum states is a von Neumann measurement. The resulting minimum-error probability is
\be
P_{\rm ME}=\frac{1}{2}(1-\tr|\Lambda|),
\ee
where $|\Lambda|=\sqrt{\Lambda^\dag\Lambda}$.

With Eq.~(\ref{Jordan1}) and Eq.~(\ref{Jordan2}), the operator $\Lambda$ can be expressed as
\be
\Lambda&=&\sum_{[\lambda],\omega}\Lambda^{[\lambda]}_\omega+\frac{\eta_2}{d_2}\sum_{[\mu],\omega}\bigg|\begin{array}{c} [\mu]\\  \omega\end{array}\bigg\ket_2\bigg\bra\begin{array}{c} [\mu]\\  \omega\end{array}\bigg|,
\ee
with
\be
\Lambda^{[\lambda]}_\omega=\frac{\eta_2}{d_2}\bigg|\begin{array}{c} [\lambda]\\  \omega\end{array}\bigg\ket_2\bigg\bra\begin{array}{c} [\lambda]\\  \omega\end{array}\bigg|-\frac{\eta_1}{d_1}\bigg|\begin{array}{c} [\lambda]\\  \omega\end{array}\bigg\ket_1\bigg\bra\begin{array}{c} [\lambda]\\  \omega\end{array}\bigg|,
\ee
where the Young diagram $[\lambda]$ can be taken for both $\rho_1$ and $\rho_2$, while $[\mu]$ for $\rho_2$ only.
The eigenvalues of $\Lambda^{[\lambda],\omega}$ can be easily obtained as
\be
\lambda^k_{\omega,+}&=&\frac{1}{2}(c_-+\sqrt{c_+^2-(c_+^2-c_-^2)O_k^2}),\non\\
\lambda^k_{\omega,-}&=&\frac{1}{2}(c_--\sqrt{c_+^2-(c_+^2-c_-^2)O_k^2}),
\ee
with $c_\pm=\eta_2/d_2\pm\eta_1/d_1$, and we have used $k$ to denote the Young diagram $[N-k,k]$. The eigenvalue spectrum of $\Lambda^{[\lambda],\omega}$ is therefore as follows,
\be
\Lambda^{[\lambda],\omega}=\lambda_{\omega,+}^k|\lambda_{\omega,+}^k\ket\bra\lambda_{\omega,+}^k|+\lambda_{\omega,-}^k|\lambda_{\omega,-}^k\ket\bra\lambda_{\omega,-}^k|,
\ee
where $|\lambda_{\omega,+}^k\ket$ and $|\lambda_{\omega,-}^k\ket$ are the eigenvectors  corresponding to the eigenvalues $\lambda_{\omega,+}^k$ and $\lambda_{\omega,-}^k$, respectively. By some algebra, one can easily know $\lambda^k_{\omega,+}\geqslant0$ and $\lambda^k_{\omega,-}\leqslant0$, so we can get
\be
\label{PE}
P_{\rm ME}&=&\frac{1}{2}\bigg(\eta_1+\frac{\eta_2d_1}{d_2}-\sum_{k=0}^{n_C}d^k\sqrt{c_+^2-(c_+^2-c_-^2)O_k^2}\bigg),\non\\
\ee
and the corresponding measurement operators read
\be
\Pi_1&=&\sum_{k=0}^{n_C}\sum_{\omega}|\lambda_{\omega,-}^k\ket\bra\lambda_{\omega,-}^k|,\non\\
\Pi_2&=&\sum_{k=0}^{n_C}\sum_{\omega}|\lambda_{\omega,+}^k\ket\bra\lambda_{\omega,+}^k|+\sum_{[\mu],\omega}\bigg|\begin{array}{c} [\mu]\\  \omega\end{array}\bigg\ket_2\bigg\bra\begin{array}{c} [\mu]\\  \omega\end{array}\bigg|.
\ee
Obviously, $P_E$ is also dependent on the dimension $n$ and the numbers of copies in systems $A$, $B$ and $C$, and for the qubit case ($n=2$), the express in Eq.~(\ref{PE}) reproduces the results in the Ref.~\cite{Sentis}. Next, we will give some special examples to show the influence of the dimension $n$ for both unambiguous discrimination and minimum-error discrimination.

\section{Some examples}\label{sec7}
In previous works, the authors have already given some examples for qubit case to show the fact that more copies in program and data systems will give lower inconclusive probability and lower minimum-error probability for unambiguous discrimination and minimum-error discrimination between the mean input states $\rho_1$ and $\rho_2$. The results also hold for qudit cases, and therefore we do not focus on this question here. In this section, we mainly provide some examples to show the influence of the dimension $n$ on both the unambiguous discrimination and minimum-error discrimination between $\rho_1$ and $\rho_2$. For convenience sake, we set $\eta_1=\eta_2=0.5$.

\begin{figure}
\begin{center}
\epsfig{figure=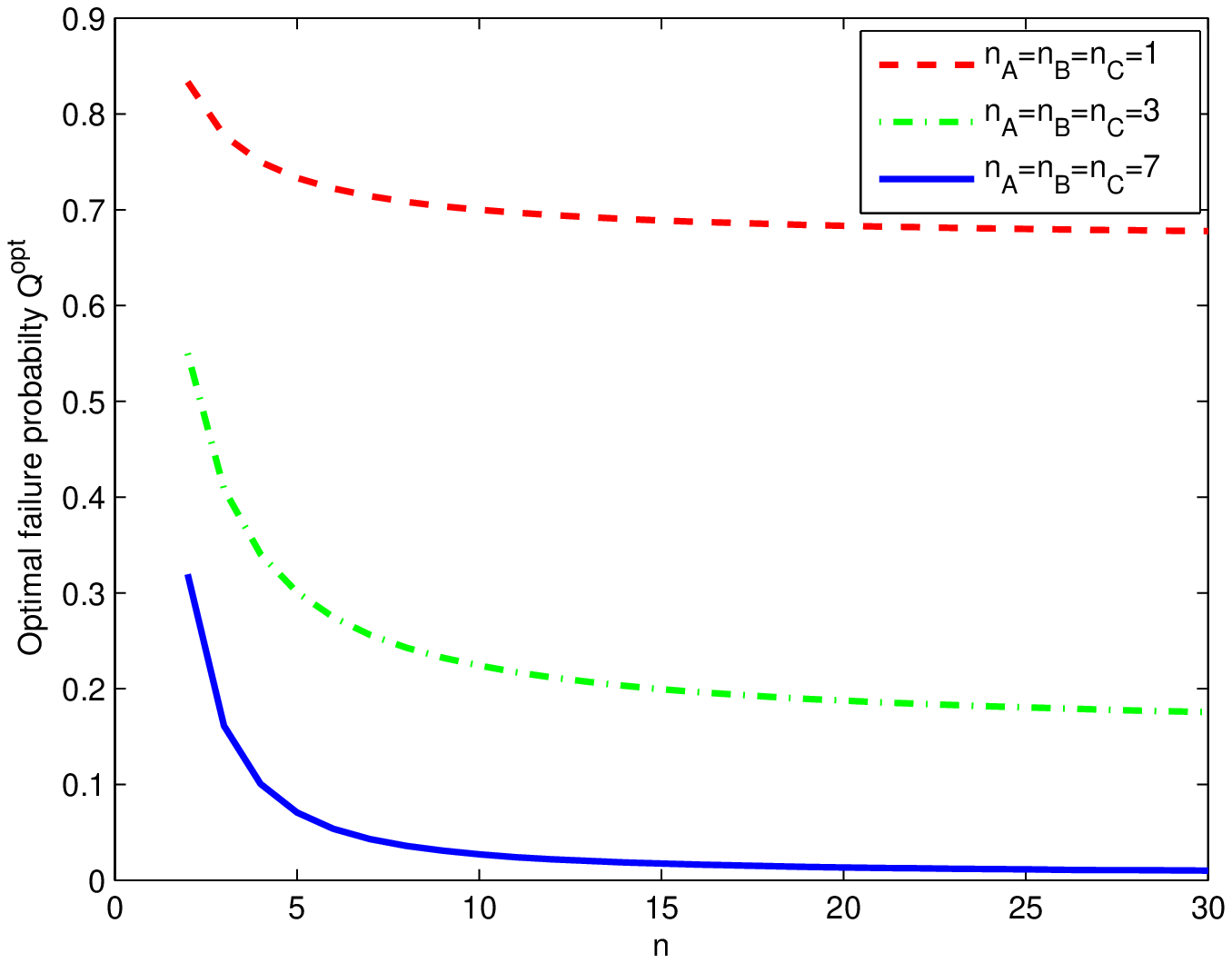,width=9.5cm}
\end{center}
\caption{(Color online) Graphs of the optimal failure probability $Q^{\rm opr}$ as a function of the dimension $n$ for $n_A=n_B=n_C=1$(dashed, red), $3$ (dotted-dashed, green) and $7$ (solid, blue).}
\label{UD}
\end{figure}
First, we consider the unambiguous discrimination between $\rho_1$ and $\rho_2$. If $n_A=n_C$, we have $d_1=d_2$, and the inequality $c_k\leqslant\eta_1\leqslant d_k$ always holds for $0\leqslant k\leqslant n_C$. Therefore, the total POVM is valid, and the the optimal inconclusive probability is reduced to
\be
Q^{\rm opt}=\frac{1}{d_1}\sum_{k=0}^{n_C}d^kO_k.
\ee 
For the cases $n_A=n_B=n_C=1,3$ and $7$, the numerical results of the failure probability $Q^{\rm opt}$ as a function of the dimension $n$ are displayed in Fig.~\ref{UD}. One can see that the optimal failure probability decreases as the dimension $n$ increases. For large $n$, there is a low bound for $Q^{\rm opt}$, and the low bound can be obtained as
\be
Q_0=\frac{\Gamma(n_A+1)\Gamma(n_B/2+1)}{\Gamma(n_A+n_B/2+1)}
\ee
for arbitrary $n_A=n_C$ and $n_B$ when $n\rightarrow\infty$. The results also show that $Q^{\rm opt}$ decreased as the number of the copies is added.

\begin{figure}
\begin{center}
\epsfig{figure=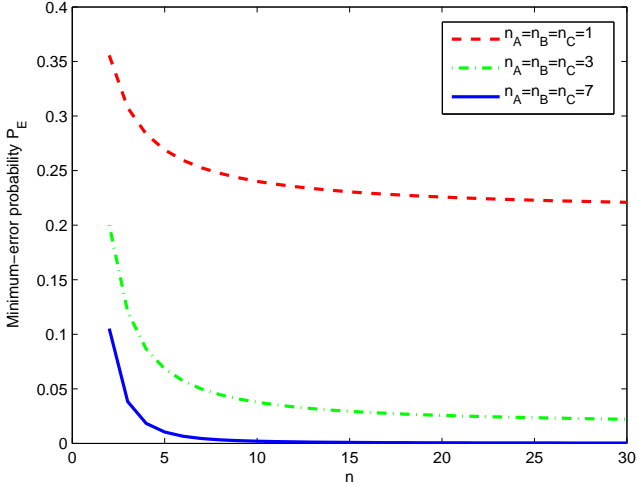,width=9.5cm}
\end{center}
\caption{(Color online) Graphs of the minimum-error probability $P_E$ as a function of the dimension $n$ for $n_A=n_B=n_C=1$(dashed, red), $3$ (dotted-dashed, green) and $7$ (solid, blue).}
\label{ME}
\end{figure}

Next, we consider the minimum-error discrimination between $\rho_1$ and $\rho_2$.  When $n_A=n_C$, Eq.~(\ref{PE}) becomes 
\be
P_{\rm ME}=\frac{1}{2}(1-\frac{1}{d_1}\sum_{k=0}^{n_C}d^k\sqrt{1-O_k^2}).
\ee
We plot the minimum-error probability versus the dimension $n$ for the cases $n_A=n_B=n_C=3,7$ and $10$ in Fig.~\ref{ME}. We see that the minimum-error probability also decreases as the dimension increases, and also decreases as the number of the copies is added. Similarly, when $n\rightarrow\infty$, the low bound for $P_{\rm ME}$ is obtained as
\be
P_0=\frac{1}{2}\bigg(1-\sum_{k=0}^{n_C}\frac{(N-2k+1)n_1!n_C!}{(N-k+1)k!(N-k)!}\sqrt{1-O_k^2}\bigg)\non\\
\ee
for arbitrary $n_A$, $n_B$ and $n_C$.

\begin{figure}
\begin{center}
\epsfig{figure=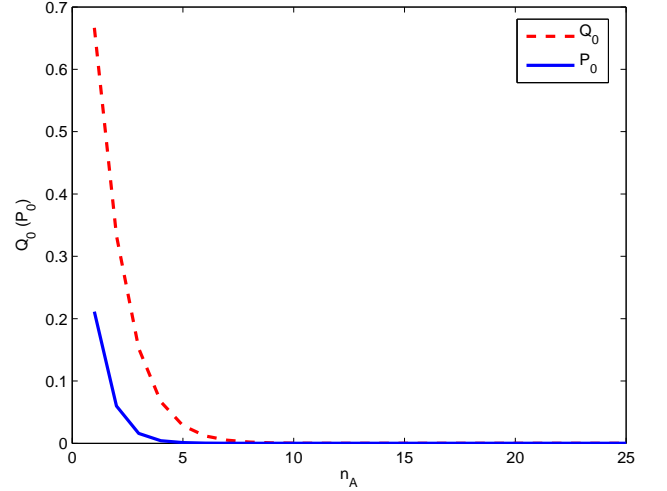,width=9.5cm}
\end{center}
\caption{(Color online) Graphs of the low bounds $Q_0$ (dashed, red) and $P_0$ (solid, blue) for $n\rightarrow\infty$ as a function of $n_A$ for the case $n_A=n_B=n_C$.}
\label{compare}
\end{figure}

For the case $n_A=n_B=n_C$, the low bounds for $Q^{\rm opt}$ and $P_{\rm ME}$ as a function of $n_A$ are depicted in Fig.~\ref{compare}. The bounds decrease as the copies are added and they both approach $0$ as $n_A\rightarrow \infty$.

A supplement material is provided together with this paper where two m-files are given. One can calculate the optimal failure probability and the minimum-error probability for the discrimination between $\rho_1$ and $\rho_2$ with arbitrary $n_A$, $n_B$ and $n_C$, and the other can plot the two probabilities versus the dimension $n$ for any ${\sl a~ priori}$  probabilities $\eta_1$ and $\eta_2$.

\section{Discussion and summary}\label{sec8}
In summary, we investigate the universal discrimination between two unknown qudit states with arbitrary numbers of copies in both data system and program systems. According to Shur's lemma, we demonstrate that the average input states are the maximally mixed states in the tensor spaces consisting of two totally symmetric spaces. The tensor spaces are reducible, and with the reducibility of $U(n)$ group, it can be decomposed into some irreducible subspaces denoted by the Young diagrams.  The Jordan bases of the mean input states are just the irreducible basis of each irreducible subspace. We also find that the inner products of the Jordan bases are determined only by the corresponding Young diagrams and thus are independent on the dimension $n$.  By the coupling theory of angular momentum, the explicit expressions of the inner products are derived. The multiplicities of the inner products are just the the dimensions of the irreducible subspaces, which can be given by the Robinson formula. 

Then, we apply the Jordan-basis method, and the problem is reduced to the discrimination between two known pure states in each two-dimensional subspace $\mathbb{H}^{[\lambda]}_\omega$. We give the optimal measurement operators for both unambiguous discrimination and minimum-error discrimination between the mixed states $\rho_1$ and $\rho_2$, where the optimal failure probability and the minimum-error probability are obtained in Eq.~(\ref{UDP}) and Eq.~(\ref{PE}), respectively. For the qubit case ($n=2$), the results in the previous works can be reproduced.

Finally, some special examples are given to show the influence of the dimension $n$ on the discrimination between $\rho_1$ and $\rho_2$. We find that both the optimal failure probability and the minimum-error probability of unambiguous discrimination and minimum-error discrimination are decreased as the dimension $n$.

\acknowledgments
The authors would like to thank Dr. Bing He for bringing the problem to them and for his many encouragements and helpful discussions. One of the authors (T.Z.) is grateful for the helpful discussions with Y. S. Li and J. X. Cui. This work was supported by the National Natural Science Foundation of China (Grants No. 10874098 and No. 11175094) and the National Basic Research Program of China (Grants No. 2009CB929402 and No. 2011CB9216002).

\appendix
\section{Partitions, Young diagrams and Young tableaux}
A partition is a way of writing a positive integer $n$ as a sum of $k$ ($k\leqslant n$) integers $\lambda_i$ satisfying
\be
n=\sum_{i=1}^k\lambda_i,\ \ \lambda_1\geqslant\lambda_2\geqslant\cdots\geqslant\lambda_k>0.
\ee
It can be pictured as a Young diagram $[\lambda]=[\lambda_1\lambda_2\cdots\lambda_k]$, which consists of $n$ cells arranged in left-justified rows, with $\lambda_i$ cells in the $i$th row. Since a partition of $n$ corresponds to a inequivalent and irreducible representation of the permutation group $S_n$,  a Young diagram $[\lambda]$ can usually be used to label an inequivalent and irreducible representation of $S_n$.

A Young tableau is an arrangement of the numbers $1,2,\cdots,n$ in a Young diagram. If the numbers increase as one moves to the right and goes down, the Young tableau is a standard Young tableau. For a given Young diagram $[\lambda]$, the number of standard Young tableaux $f^{[\lambda]}$ is equal to the dimension of the irreducible representation $[\lambda]$ of the permutation group $S_n$, and can be calculated by the formula
\be
\label{dimf}
f^{[\lambda]}=\frac{n!}{\prod_{ij}g_{ij}},
\ee
where $g_{ij}$ is the hook length for the cell in the $i$th row and $j$th column of the Young diagram $[\lambda]$. A hook of a cell consists of this given cell together with all those to the right in the same row and lower in the same column, and the number of cells in the hook is called the hook length. The $m$-th standard Young tableau can be denoted by $T^{[\lambda]}_m$ for a given Young diagram $[\lambda]$, where $m=1,2,\cdots,f^{[\lambda]}$.

\section{Representation theory of $U(n)$ group}
The tensor product $U^{\otimes k}$ is a faithfull representation of the $n$-dimensional unitary group $U(n)$ on the tensor space $\mathcal{H}^{\otimes k}$, and it is reducible. Therefore, it can be decomposed into~\cite{Sunbook}
\be
U^{\otimes k}=\bigoplus_{[\lambda],m}U^{[\lambda]}_m,
\ee
and correspondingly, $\mathcal{H}^{\otimes k}$ can be decomposed into
\be
\mathcal{H}^{\otimes k}=\bigoplus_{[\lambda],m}\mathcal{H}^{[\lambda]}_m.
\ee
Here, $U^{[\lambda]}_m$ is an irreducible representation of $U(n)$ group on the subspace $\mathcal{H}^{[\lambda]}_m$, corresponding to a standard Young tableau $T^{[\lambda]}_m$, where the number of rows in $[\lambda]$ is no more than $n$. For $[\lambda]=[k]$ or $[\lambda]=[1^k]$, $m$ can take only one value, and we can omit the label $m$ in these cases.

The irreducible subspaces $\mathcal{H}^{[\lambda]}_m$ can be constructed via a standard way,
\be
\mathcal{H}^{[\lambda]}_m=O^{[\lambda]}_{mm}\mathcal{H}^{\otimes k},
\ee
where the projector operator $O^{[\lambda]}_{mm}$ are the orthogonal units of the permutation group $S_k$. For the details of $O^{[\lambda]}_{mm}$, see Ref.~\cite{Sunbook}.
If $\{e_i\}$ ($i=1,2,\cdots,n$) form the complete orthogonal basis of $\mathcal{H}$, the complete orthogonal bases of the irreducible space $\mathcal{H}^{[\lambda]}_m$ can be constructed by $O^{[\lambda]}_{mm}$, and a basis vector of $\mathcal{H}^{[\lambda]}_m$ can be obtained as
\be
\xi^{[\lambda]}_{m,i_1i_2\cdots i_k}=O^{[\lambda]}_{mm}e_{i_1i_2\cdots i_k},
\ee
where $e_{i_1i_2\cdots i_k}=e_{i_1}\otimes e_{i_2}\otimes\cdots\otimes e_{i_k}$ ($i_1,i_2,\cdots,i_k=1,2,\cdots,n$) are the complete orthogonal basis of $\mathcal{H}^{\otimes k}$. However, the bases $\{\xi^{[\lambda]}_{m,i_1i_2\cdots i_k}\}$ are linearly dependent, and to give the independent basis of $\mathcal{H}^{[\lambda]}_m$, we should first introduce the concept of Weyl tableaux.

A Weyl tableau is a Young diagram $[\lambda]$ whose cells are filled with some of the numbers $1,2,\cdots,k$ under the restrictions that\\
(a) the numbers do not decrease in the row as one moves to the right;\\
(b) the numbers increase in the column as one goes down.

If the index $i_1,i_2,\cdots,i_k$ take the values in the Wely tableaux, $\xi^{[\lambda]}_{m,i_1i_2\cdots i_k}$ are independent and form a complete orthogonal bases of $\mathcal{H}^{[\lambda]}_m$. Thus, the dimension of subspace $\mathcal{H}^{[\lambda]}_m$ is the number of the Weyl tableaux for $[\lambda]$, which has been given by the Robinson formula
\be
\label{dimd}
d^{[\lambda]}=\prod_{ij}\frac{n-i+j}{g_{ij}},
\ee
where $g_{ij}$ is the hook length. For convenience sake, we use $\bigg|\begin{array}{c} [\lambda] \\
m, \omega\end{array}\bigg\ket$ to denote the \emph{normalized} basis vectors for $\mathcal{H}^{[\lambda]}_m$, where $m$ and $\omega$ correspond to the standard Young tableaux and Weyl tableaux, respectively, $m=1,2,\cdots,f^{[\lambda]},\omega=1,2,\cdots,d^{[\lambda]}$. Furthermore, $\bigg|\begin{array}{c} [\lambda] \\
m, \omega\end{array}\bigg\ket$ are the standard basis of $S_k$ and the irreducible basis of $U(n)$~\cite{Chenbook}.

Suppose $U^{[\lambda]}$ and $U^{[\mu]}$ are two irreducible representation for the unitary group $U(n)$, and the tensor product $U^{[\lambda]}\otimes U^{[\mu]}$ is also a representation of $U(n)$, but usually reducible. With Littlewood rule, $U^{[\lambda]}\otimes U^{[\mu]}$ can be decomposed into
\be
U^{[\lambda]}\otimes U^{[\mu]}=\bigoplus_{[\sigma]}\{[\lambda][\mu][\sigma]\}U^{[\sigma]},
\ee
with $\{[\lambda][\mu][\sigma]\}$ the multiplicity for $[\sigma]$. One should notice that the irreducible basses for $U^{[\sigma]}$ are usually not the standard basis of permutation group.

\end{document}